\pdfoutput=1
\documentclass[prl,aps,showpacs,twocolumn,a4]{revtex4}
\usepackage{amsmath,amssymb}
\usepackage{graphicx}
\usepackage{psfrag}
\usepackage{array}
\usepackage{epsfig}
\usepackage{subfigure}
\usepackage[usenames]{color}
\newcommand{\distance}{x}
\newcommand{\lp}{\ell_\text{p}}

\newcommand{\kT}{k_\text{B}T}

\newcommand{\e}{\mbox{e}}

\newcommand {\be}  {\begin{equation}}
\newcommand {\ee}  {\end{equation}}
\newcommand {\bea}  {\begin{eqnarray}}
\newcommand {\eea}  {\end{eqnarray}}

\begin{document}

\title{Velocity oscillations in actin-based motility} \author{Azam
  Gholami$^1$, Martin Falcke$^1$, Erwin Frey$^{2}$ }
\affiliation{$^{1}$Hahn-Meitner-Institute, Dept. Theoretical
Physics, Glienicker
  Str. 100, 14109 Berlin, Germany \\
  $^2$Arnold Sommerfeld Center for Theoretical Physics and Center of
  NanoScience, Ludwig-Maximilians-Universit\"{a}t, Theresienstr. 37,
  80333 M\"{u}nchen, Germany}

\date{\today}
\begin{abstract}
  We present a simple and generic theoretical description of
  actin-based motility, where polymerization of filaments maintains
  propulsion. The dynamics is driven by polymerization kinetics at the
  filaments' free ends, crosslinking of the actin network,
  attachment and detachment of filaments to the obstacle
  interfaces and entropic forces. We show that spontaneous
  oscillations in the velocity emerge in a broad range of parameter values,
  and compare our findings with experiments.
\end{abstract}
\maketitle

Force generation by semiflexible polymers is versatilely used for cell
motility. The leading edge of lamellipodia of crawling cells
\cite{Bray-2001} is pushed forward by a polymerizing actin network and
bacteria move inside cells by riding on a comet tail of growing actin
filaments \cite{Plastino-Sykes-2005,Cossart2005}. In vivo systems are
complemented by in vitro assays using plastic beads and lipid vesicles
\cite{Loisel-Carlier-1999,Marcy-Prost-2004,Parekh-Theriot-2005}. The
defining feature of semiflexible polymers is the order of magnitude of
their bending energy which is in the range of $\kT$. They undergo
thermal shape fluctuations and the force exerted by the filaments
against an obstacle arises from \emph{elastic and entropic}
contributions \cite{mogilner-oster:96a,gholami-wilhelm-frey:2006}.

Mathematical models have quantified the force generated by actin
filaments growing against obstacles
\cite{Hill-1981,mogilner-oster:96a,gholami-wilhelm-frey:2006}. The
resisting force depends on the obstacle which is pushed. In case
of pathogens, it has a small component from viscous drag of the
moving obstacle but consists mainly of the force exerted by actin
filaments bound to the surface of the bacteria and pulling it
backwards~\cite{Cameron-Theriot-2001,Kuo-McGrath-2000}. The
tethered ratchet model \cite{mogilner-oster-2003} is a
mathematical formulation of these experimental findings in terms
of the dynamics of the number of attached and detached polymers.
The starting point of our approach will be the dynamics of the
distributions of the free length of both polymer populations.

\begin{figure}[htbp]
 \includegraphics[width=\columnwidth] {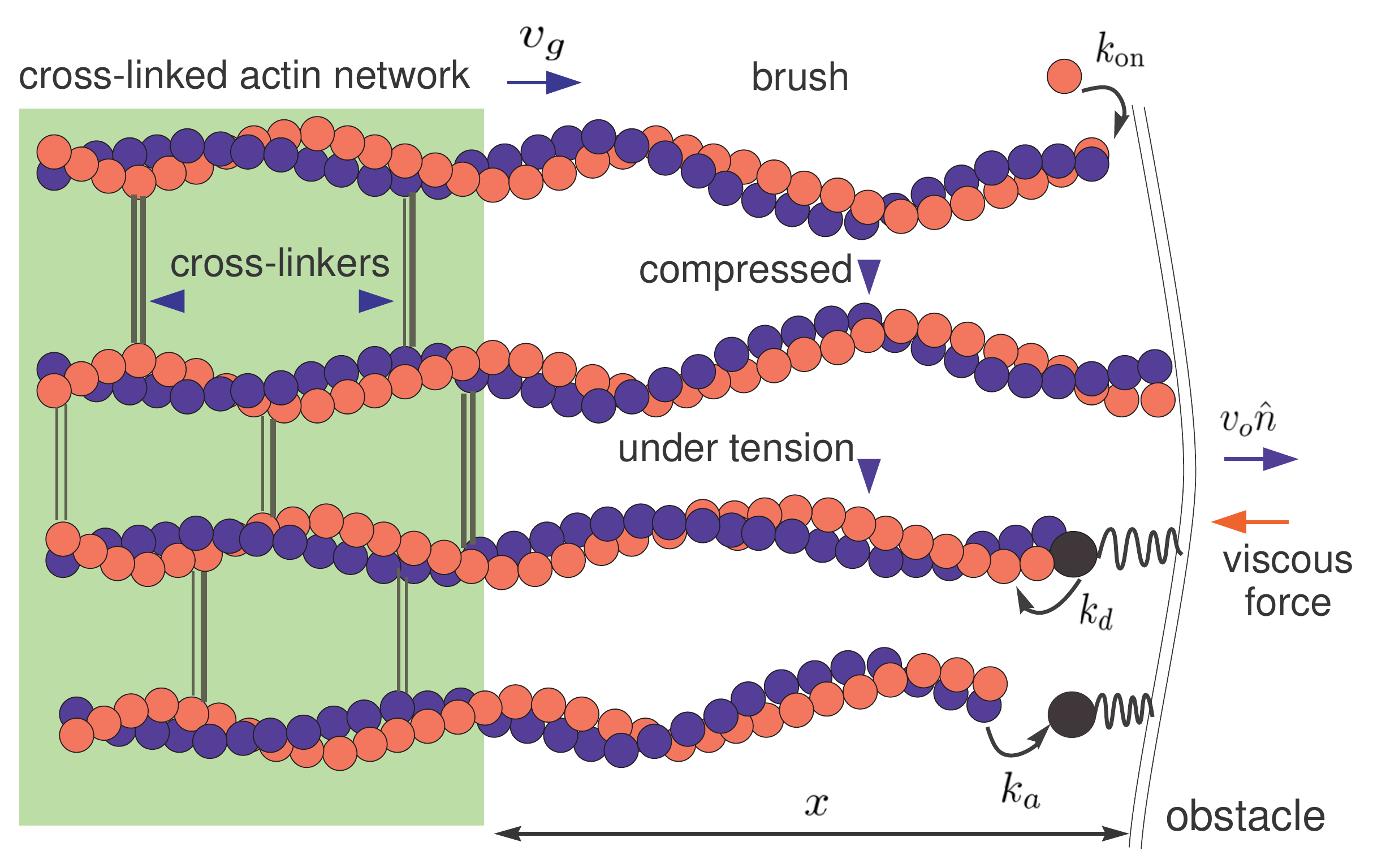}
 \caption{(color online) Schematic representation of an ensemble of actin
   filaments oriented at $\vartheta=0$ with respect to the normal
   $\hat n$ of an obstacle interface, which may either be a cell
   membrane or a bacterium. While attached filaments are under
   tension and pull the interface back, detached filaments are
   compressed, elongate by polymerization with rate $k_\text{on}$ and
   push the interface forward.  All filaments in the brush are firmly
   anchored in a cross-linked network, whose front advances with
   velocity $v_g$ reducing the free length $l$ of the filaments.
   Attached filaments detach with stress dependent rate $k_d$ and
   detached filaments attach with constant rate $k_a$.  $v_o$ is the
   interface velocity in the extracellular medium, and $\distance$ is
   the distance between the front of the network and the interface.}
 \label{fig:setup}
\end{figure}

Actin polymerization in living cells and extracts is controlled by
a complex molecular network \cite{Cossart2005}. Nucleation of new
filaments, capping of existing ones, exchange of ADP for ATP on
actin monomers, buffering of monomers etc. all contribute to that
control and have been modeled
\cite{Carlsson2003,mogilner-oster-2003,Othmer}. Our goal is not to
model the full complexity of that biochemical network. Rather we
focus on the core process of force generation and force balance
ensuing from the interplay between bound pulling filaments and
polymerizing pushing filaments, the transition between these two
groups and the motion of the whole force generating configuration.
This is motivated by recent observations of complex dynamics in
simple reconstituted systems: the velocity of beads or pathogens
propelled by actin polymerization may oscillate
\cite{Cossart97,Prost2000,Prost2005}. Our goal is to describe the
dynamics of such biochemically simpler systems and find a robust
microscopic description for oscillation mechanisms, which may then
be controlled by higher order processes. Such a study is meant to
complement investigations based on a continuum
approach~\cite{Prost2000,Prost2005}.

We consider a fixed number $N$ of actin
filaments~\cite{Brieher2004} firmly anchored into a rigid
cross-linked network, which advances with velocity $v_g$; for an
illustration see Fig.~\ref{fig:setup}. Filaments of variable
length $l$ are either attached to the obstacle interface via a
protein complex or detached from it, with time-dependent number
distributions denoted by $N_a(l,t)$ and $N_d(l,t)$, respectively.
In the detached state, filaments polymerize at a velocity $v_p
(l,\distance)$, which depends on both the polymer length $l$ and
the distance $\distance$ between rigid support and obstacle.
Transitions between the two filament populations occur with a
constant attachment rate $k_a$ and a stress-dependent detachment
rate $k_d$ \cite{Evans-Ritchie-1999}. This results in the
evolution equations
\begin{subequations}
\begin{align}
  \frac{\partial}{\partial t} N_d- \frac{\partial}{\partial l}
  \left[\frac{l}{x} v_g(l) - v_p\right] N_d&=-k_a~N_d+k_d~N_a\, ,
\label{Eqn:Nd_TimeEvolution_General}\\
\frac{\partial}{\partial t} N_a - \frac{\partial}{\partial l}
\left[\frac{l}{x} v_g (l)~~~~~~\right]N_a&=~~k_a~N_d-k_d~N_a\, .
\label{Eqn:Na_TimeEvolution_General}
\end{align}
\label{Eqn:TimeEvolution_General}
\end{subequations}
\noindent The right hand side of
Eq.~\ref{Eqn:TimeEvolution_General} describes attachment and
detachment process. The second term on the left hand side accounts
for the gain and loss of attached and detached polymers due to the
dynamics of the polymer mesh, growing with velocity $v_g$, and the
polymerization kinetics of the filaments in the brush. The
correction factor $l/x$ in front of $v_g$ is due to the fact that
for bent polymers the rigid network swallows by this amount more
in contour length than for straight filaments. This factor is
equal to $1$ for $l \leq \distance$.

Processes contributing to the growth of the rigid polymer mesh are
entanglement and crosslinking of filaments in the brush. Both imply a
vanishing $v_g$ for $l \to 0$, since short polymers do not entangle
and crosslinking proteins are unlikley to bind to them.  At the same
time $v_g$ can not grow without bound but must saturate at some value
$v_g^\text{max}$ due to rate limitations for crosslinker binding. This
suggests to take the following sigmoidal form
\begin{equation}
   v_g (l) = v_g^\text{max} \tanh (l/\bar l) \, ,
\label{Eqn_vg}
\end{equation}
with a characteristic length scale $\bar l$.

The polymerization rate is proportional to the probability of a
gap of sufficient size $d$ ($\approx 2.7$~nm) between the polymer
tip and the obstacle for insertion of an actin monomer
\cite{mogilner-oster:96a}. This implies an exponential dependence
of $v_p$ on the force $F_\text{d}$ by which the polymer pushes
against the obstacle,
\begin{equation}
\label{Eqn:Vp_ForceDependence}
  v_p(l,\distance) =  v_p^\text{max}
                  \exp \left[ -d \cdot F_\text{d}(l,\distance)/\kT
                  \right] \, .
\end{equation}
Here, $v_p^\text{max} \approx
500$~nm~s$^{-1}$~\cite{mogilner-oster:96a} is the free
polymerization velocity.  For the entropic force $F_\text{d}$ we
use the results obtained in Ref.~\cite{gholami-wilhelm-frey:2006}
for $D=2, \, 3$ spatial dimensions, where we take the accepted
value of $\lp\approx 15~\mu$m~\cite{persistence_length} for the
persistence length of F-actin.

The dynamics of the distance $\distance$ between grafted end of the
filament and the obstacle interface (see Fig.~\ref{fig:setup}) is
given by the difference of the average $v_g$ and the velocity of the
obstacle
\begin{eqnarray}
 \label{Eqn:Zeta_TimeEvolution_General}
&&\partial_t \distance
   =-\frac{1}{N}\int_0^\infty dl~v_g(l)~
    \left[ N_a(l,t)+N_d(l,t) \right]\\
&&+\frac{1}{\zeta}\int_0^\infty
   dl~\left[ N_a(l,t)~F_a (l,\distance)+N_d(l,t)
   ~F_\text{d}(l,\distance) \right]
\nonumber\, ,
\end{eqnarray}
where $\zeta$ is an effective friction coefficient of the
obstacle. The force $F_a(l,\distance)$ acting on the obstacle
interface results from the compliance of the filaments attached to
it by some linker protein complex, which we model as springs with
spring constant $k_l$ and zero equilibrium length. This complex
has a nonlinear force-extension relation which we approximate by a
piece-wise linear function; for details see the supplementary
material. Let $R_\parallel \approx l [ 1-l(D-1)/4\lp ]$ be the
equilibrium length of the filament.  Then, the elastic response of
filaments experiencing small compressional forces ($\distance \le
R_\parallel$) is approximated by a spring constant $k_{\parallel}
= 12 \kT~\lp^2/(D-1) l^4$~\cite{kroy-frey:96}. For small pulling
forces ($\distance\ge R_{\parallel}$), the linker-filament complex
acts like a spring with an effective constant $k_\text{eff}=k_l
k_\parallel/(k_l+k_\parallel)$. In the strong force regime, the
force-extension relation of the filament is highly nonlinear and
diverges close to full stretching~\cite{Marko-Siggia-95}.
Therefore, only the linker will stretch out.  The complete
force-extension relation is captured by
\begin{equation}
F_a =
\begin{cases}
  -k_{\parallel}(\distance-R_\parallel) \, ,&
  \text{$\distance\leq R_\parallel$ }  ,\\
  -k_{\text{eff}} (\distance-R_\parallel) \, ,&
  \text{$R_\parallel < \distance < l$ }  ,\\
  -k_l (\distance-l)-k_{\text{eff}} (l-R_\parallel) \, ,&
  \text{$\distance\ge l$  .}
\end{cases}
\end{equation}

Finally, we specify the force-dependence of the detachment rate by
\begin{equation}
\label{Eqn:Kd_StressDependence}
 k_d = k_d^0~\exp\left[-d \cdot F_a (l,\distance)/k_\text{B}T \right]\, .
\end{equation}
Here, $k_d^0\approx 0.5~\text{s}^{-1}$~\cite{mogilner-oster-2003} is
the detachment rate in the absence of forces and we have followed
Ref.~\cite{Evans-Ritchie-1999}.

Eq.~\ref{Eqn:Nd_TimeEvolution_General} has a singularity at
$v_p(l_s)=v_g(l_s) l_s/\distance$ since the coefficient of the
derivative of $N_d$ with respect to $l$ is zero at $l_s$. To
illustrate the key physical features at that singularity, we start
with the simple equation $\partial_t N_d- \partial_l [v_g(l) l /
\distance - v_p(l,\distance)] N_d=0$ with $\distance$ kept
constant. Then those parts of the distribution of $N_d$ with
$l<l_s$ will grow and catch up with $l_s$ since $v_g(l) l /
\distance - v_p(l,\distance)$ is positive there, while the parts
with $l>l_s$ will shorten towards $l_s$. As a consequence the
whole distribution will become concentrated at $l_s$. To quantify
this heuristic argument we expand $v_g(l) l / \distance -
v_p(l,\distance)$ up to linear order around $l_s$ like
$v_1(l-l_s)$ and use the method of characteristics to solve the
equation. Starting initially with a Gaussian distribution we
obtain $N_d(l,t)=c(t)\exp[-(l-\bar l(t))^2/w(t)^2]$ with
$c(t)=c_0\exp(v_1t)$, $\bar l(t) = l_s + (\bar
l_0-l_s)\exp(-v_1t)$ and $w(t)=w_0\exp(-v_1t)$. This shows that
$N_d$ evolves to a monodisperse distribution which is localized
around $l_s$. Its width decreases exponentially with time while
its height grows exponentially. The time scale for this
contraction is given by $[\partial_l(v_g l / \distance -
v_p)]^{-1}$.

Since the same kind of singularity also occurs in the full set of
dynamic equations, Eqs.~\ref{Eqn:TimeEvolution_General}, we may
readily infer that $N_a$ and $N_d$ evolve into delta-functions with
that dynamics.  This is well supported by simulations, and allows us
to continue with the ansatz
\begin{subequations}
\begin{align}
N_d(l,t) = n_d(t)~\delta(l-l_d(t)) \, ,
\label{Eqn:Nd_DeltaAssumption}\\
N_a(l,t) = n_a(t)~\delta(l-l_a(t)) \, .
\label{Eqn:Na_DeltaAssumption}\end{align}
\label{Eqn:DeltaAssumption}
\end{subequations}
It defines the dynamic variables $n_d(t)$, $l_d(t)$, $n_a(t)$ and
$l_a(t)$. Upon inserting Eqs.~\ref{Eqn:DeltaAssumption} into
Eqs.~\ref{Eqn:TimeEvolution_General} and
Eq.~\ref{Eqn:Zeta_TimeEvolution_General}, we obtain the following set
of ordinary differential equations
\begin{subequations}\label{Eqn:Eight}
\begin{align}
  \partial_t
  l_d(t)&=~v_p(l_d,\distance)-\frac{l_d}{\distance}v_g(l_d)+k_d\frac{n_a}{n_d}(l_a-l_d)
  \, ,\label{Eqn:ld_TimeEvolution_Simplified}\\
  \partial_t l_a(t) &=
  -\frac{l_a}{\distance}v_g(l_a)+k_a\frac{n_d}{n_a}(l_d-l_a)
  \, ,\label{Eqn:la_TimeEvolution_Simplified}\\
  \partial_t n_a(t) &=-k_d(l_a,\distance)~n_a(t)+k_a~n_d(t)
  \, ,\label{Eqn:na_TimeEvolution_Simplified}\\
  \partial_t \distance(t) &=\frac{1}{\zeta} \left[
    n_a(t)~F_\text{a}(l_a,\distance)+n_d(t)~ F_\text{d}(l_d,\distance)
  \right]
  \nonumber\\
  &-\frac{1}{N}[v_g(l_a)~n_a(t)+v_g(l_d)~n_d(t)] \, ,
  \label{Eqn:Zeta_TimeEvolution_Simplified}
\end{align}
\end{subequations}
where $n_d(t) =N-n_a(t)$ since we keep the total number of filaments
fixed.

The values of many parameters in the dynamics can be estimated using
known properties of actin filaments. We choose the linker spring
constant $k_l\approx 1$ pN nm$^{-1}$ \cite{mogilner-oster-2003} and
assume $N=200$ \cite{mogilner-oster-2003} filaments to be crowded
behind the obstacle. A realistic value of the drag coefficient
$\zeta$ is $10^{-3}~$pN~s~nm$^{-1}$ but results did not change
qualitatively for a range from $10^{-5}~$pN~s~nm$^{-1}$ to
$1~$pN~s~nm$^{-1}$.

\begin{figure}[htbp]
  \includegraphics[width=\columnwidth]{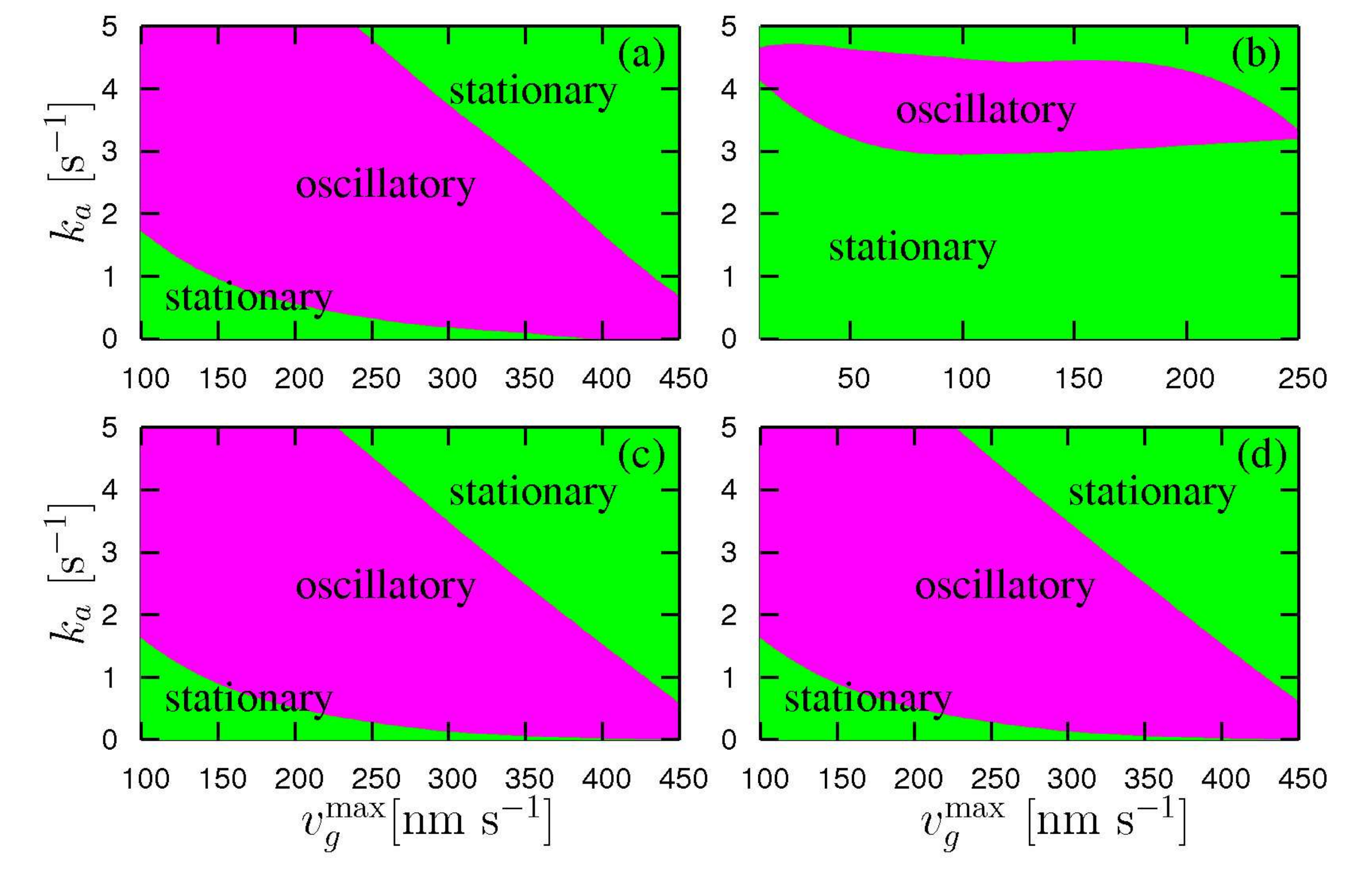}
\caption{(color online) Phase diagram of
  Eqs.~\ref{Eqn:ld_TimeEvolution_Simplified} -
  \ref{Eqn:Zeta_TimeEvolution_Simplified} outlining stationary and
  oscillatory regimes with $\zeta = 10^{-3}$ pN~s~nm$^{-1}$ for (a-c)
  and (a) $D=2$, $\vartheta = 0$ , (b) $D=2$, $\vartheta = \pi/4$, (c)
  $D=3$, $\vartheta = 0$ and (d) $D=3$, $\vartheta = 0$, $\zeta =
  10^{-5}$ pN~s~nm$^{-1}$. $\bar l$=100~nm, all other parameter
  values are specified in the text.} \label{fig:PhaseDiagram}
\end{figure}

We have numerically solved Eqs.~\ref{Eqn:Eight} in both $D=2$ and
$D=3$ dimensions, and found the dynamic regimes shown in
Fig.~\ref{fig:PhaseDiagram}: \emph{stationary states and
  oscillations}. The existence of an oscillatory regime is very robust
against changes of parameters within reasonable limits including
the spatial dimension.  We checked robustness against changes in
the parameter values for the number of polymers $N$, $\bar l$ (see
Eq.~\ref{Eqn_vg}), $k_l$, $v_p^\text{max}$ and $k_d^0$, in
addition to the examples shown in Fig.~\ref{fig:PhaseDiagram}. In
general, we find that oscillations occur for
$v_g^\text{max}\lesssim 500$ nm~s$^{-1}$ and within a range of
values for $k_a$. Note that the oscillatory region in parameter
space depends on the orientation $\vartheta$ of filaments with
respect to the obstacle surface, i.e. oscillating and
non-oscillating sub-populations of filaments may coexist in the
same network.
\begin{figure}[htbp]
  \centering {\label{fig4a:Oscillation}
    \includegraphics[width=0.5\columnwidth]{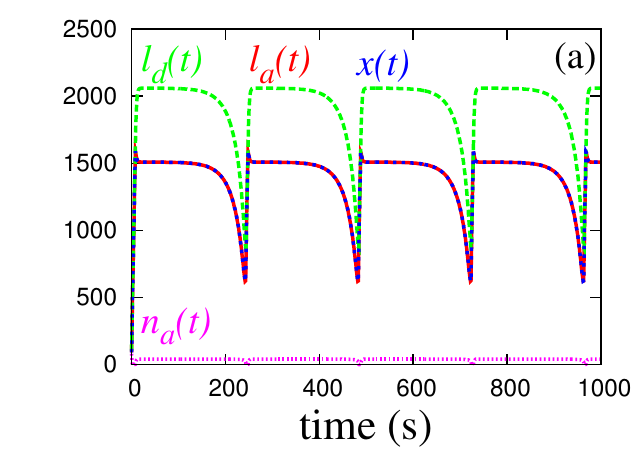}}
  \hspace{-0.5truecm} {\label{fig4b:Oscillation}
    \includegraphics[width=0.5\columnwidth]{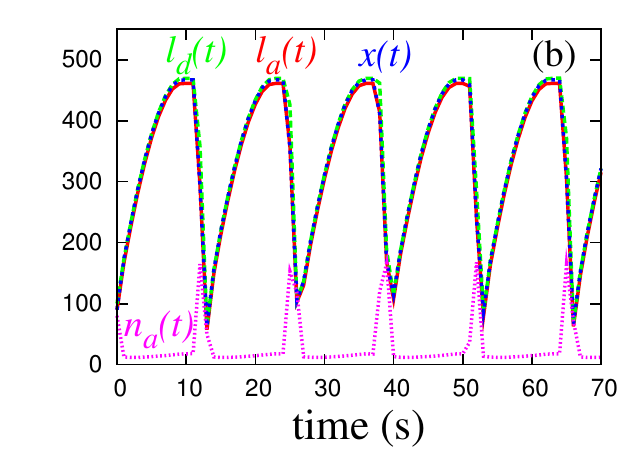}}
\caption{(color online) $\distance$ , $l_a$ , $l_d$ (in nm) and
  $n_a$ as a function of time, as obtained from a numerical
  solutions of Eqs.~\ref{Eqn:ld_TimeEvolution_Simplified} -
  \ref{Eqn:Zeta_TimeEvolution_Simplified} with $v_g^\text{max} =
  300~\text{nm}~\text{s}^{-1}$ and (a) $k_a = 0.143~\text{s}^{-1}$ (b)
  $k_a=3.49~\text{s}^{-1}$. $D=3$, $\bar l$=100~nm in both panels.}
\label{fig4:Oscillation}
\end{figure}

Oscillations appear with finite amplitude and period at the lower
boundary of the oscillatory region; compare the example shown in
Fig.~\ref{fig4:Oscillation}a. The stationary state changes stability
slightly inside the oscillatory regime and oscillations set in with a
finite period. That is compatible with oscillations appearing by a
saddle node bifurcation of limit cycles. The upper boundary of the
oscillatory region is determined by a Hopf bifurcation. An example of
an oscillation close to that bifurcation is shown in
Fig.~\ref{fig4:Oscillation}b. More details on the phase diagram will
be published elsewhere~\cite{Gholami-Falcke-Frey}.

We start with the description of oscillations in the phase with
$v_g > v_p$, i.e., decreasing lengths $\distance$, $l_a$ and
$l_d$; see Fig.~\ref{fig4:Oscillation}. Then the magnitude of
pulling and pushing forces increases due to their
length-dependence. When the pushing force becomes too strong, an
avalanche-like detachment of attached filaments is triggered and
the obstacle jerks forward; compare the steep rise in $l_d$, $l_a$
and $x$ shown in Fig.~\ref{fig4:Oscillation}. That causes a just
as sudden drop of the pushing force.  With low pushing force now,
polymerization accelerates and increases the length of detached
filaments. The restoring force of attached filaments is weak in
this phase due to their small number. Hence, despite of not so
strong pushing forces, the obstacle moves forward.  In the
meantime, some detached filaments attach to the surface such that
the average length and number of attached filaments increases as
well. When the detached filaments are long enough to notice the
presence of the obstacle interface, they start to buckle. This, in
turn, increases the pushing force and slows down the
polymerization velocity.  Therefore, the graft velocity now
exceeds the polymerization velocity and the average lengths of
attached and detached filaments start to decrease again and the
cycle starts anew. The period of oscillations is dependent on the
parameter values. It reduces from $240$~s in
Fig.~\ref{fig4:Oscillation}a to $13$~s in
Fig.~\ref{fig4:Oscillation}b as $k_a$ increases from
$0.143~\text{s}^{-1}$ to $3.49~\text{s}^{-1}$ at
$v_g^\text{max}=300~\text{nm}~\text{s}^{-1}$.

\begin{figure}[htbp]
  \centering {\label{fig4a:Oscillation}
    \includegraphics[width=0.5\columnwidth]{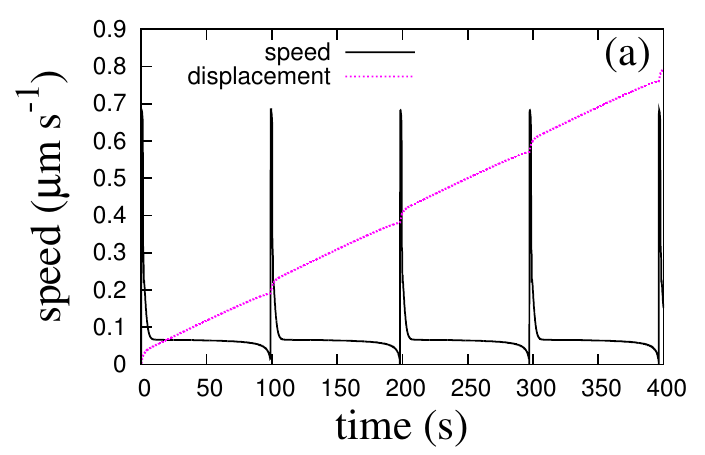}}
  \hspace{-0.65cm} {\label{fig4b:Oscillation}
    \includegraphics[width=0.5\columnwidth]{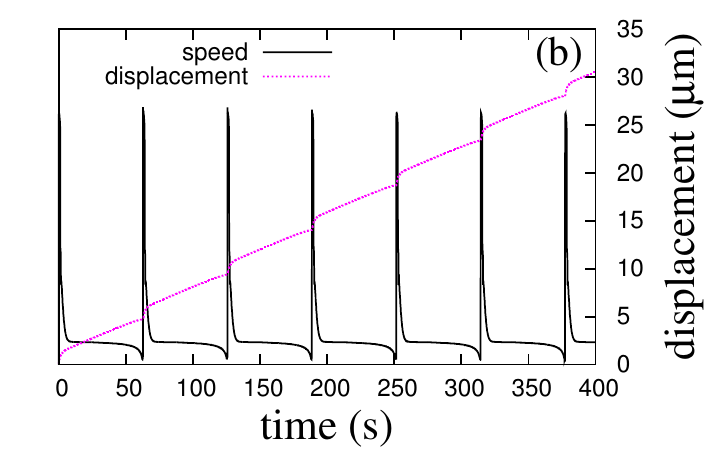}}
\caption{(color online) Velocity and displacement of the obstacle
  as a function of time with (a) $k_a
  = 0.9~\text{s}^{-1}$, (b) $k_a = 1~\text{s}^{-1}$. $v_p^\text{max} =
  750~\text{nm}~\text{s}^{-1}$, $v_g^\text{max} =
  75~\text{nm}~\text{s}^{-1}$, $k_d^0 = 0.1~\text{s}^{-1}$, $\bar l$=100~nm
  $\zeta = 10^{-3}$ pN~s~nm$^{-1}$ and $D=3$ in both panels.}
\label{Speed-Displacement}
\end{figure}

The oscillations in $\distance$ correspond to the saltatory motion
of the obstacle in the lab frame and the oscillations of its
velocity since $v_g$ stays essentially constant. An illustration
is shown in Fig.~\ref{Speed-Displacement} for a given set of
parameters which leads to oscillations with periods of the order
of $100~\text{s}$ and velocity of the order of $0.7~\mu
\text{m}~\text{s}^{-1}$. This is in good agreement with the
results of experiments on oscillatory \emph{Listeria}
propulsion~\cite{Cossart97}. The period of velocity oscillations
with beads propelled by actin polymerization differs from those of
\emph{Listeria} by one order of magnitude ($8-15$~min
\cite{Prost2005}). Periods of that length can be obtained within
our model upon using values for $k_a$ close to the lower boundary
of the oscillatory regime.

We have also studied the system when the network is oriented at an
angle $\vartheta=\pi/4$. In this case, the spring constant of the
attached filaments parallel to $\hat n$ for $D=2$ reads
$k_\parallel^{-1}(\vartheta) = 4\lp^2 [ \frac{\epsilon}{2} +
\e^{-\epsilon/2} - 1 + \cos 2\vartheta ( \frac{1}{4} +
\frac{1}{12} \e^{-2\epsilon} - \frac{1}{3} \e^{-\epsilon/2}) -
\cos^2\vartheta (\e^{-\epsilon/2} - 1)^2] / \kT$, where $\epsilon
= l/l_p$ and $R_\parallel(\vartheta) = l(1-l/4\lp) \cos\vartheta$
~\cite{kroy-frey:96}. For the pushing force of a filament grafted
at $\vartheta = \pi/4$, we use the results of the factorization
approximation given in Ref.~\cite{gholami-wilhelm-frey:2006},
which is well valid for a stiff filament like actin. A numerical
solution of
Eqs.~\ref{Eqn:ld_TimeEvolution_Simplified}-\ref{Eqn:Zeta_TimeEvolution_Simplified}
results in the phase diagram shown in
Fig.~\ref{fig:PhaseDiagram}(b) with the adapted forms of
$F_\text{d}$ and $F_\text{a}$. The main effect is that one needs
higher values for the attachment rates and lower values for $v_g$
to obtain oscillations.

In summary, we have presented a simple and generic theoretical
description of oscillations arising from the interplay of
polymerization driven pushing forces and pulling forces due to
binding of actin filaments to the obstacle. The physical mechanism
for such oscillations relies on the load-dependence of the
detachment rate and the polymerization velocity, mechanical
restoring forces and eventually also on the cross-linkage and/or
entanglement of the filament network. The oscillations are very
robust with respect to changes in various parameters, i.e. are
generic in this model. Therefore, complex biochemical regulatory
systems supplementing the core process described here may rather
stabilize motion and suppress oscillations than generate them.

Oscillations of the velocity were observed during propulsion of
pathogens by actin polymerization. There, the core mechanism
described here is embedded into a more complex control of
polymerization, which e.g.  also comprises nucleation of new
filaments and capping of existing ones. Hence, the study presented
here can not be expected to fully capture all features of such
processes. Our results still agree well with respect to velocity
spike amplitudes and periods in \emph{Listeria} experiments
reported in Refs.~\cite{Cossart97,Prost2000}. The velocity in
between spikes appears to be smaller in experiments than in our
simulations.  This may be accounted for in our model by including
capping of filaments upon dissociation from the obstacle. Periods
may also become longer when capping and nucleation were included
since it would take longer to restore the pushing force after the
avalanche like rupture of attached filaments. Altogether,
qualitative and quantitative comparison with experiments suggests
that our model may be a promising candidate for a robust mechanism
of velocity oscillations in actin-based bacteria propulsion.

We thank R. Straube and V. Casagrande for inspiring discussions.
E.F. acknowledges financial support of the German Excellence
Initiative via the program ''Nanosystems Initiative Munich
(NIM)''. A.G. acknowledges financial support of the IRTG
''Genomics and Systems Biology of Molecular Networks'' of the
German Research Foundation.

\end{document}